\documentclass[useAMS,usenatbib,referee]{mn2e}

\usepackage{amsmath}
\usepackage{amssymb}
\usepackage{graphicx}
\usepackage{dcolumn}
\usepackage{bm}
\usepackage{ulem}

\title[Electrical and thermal conductivity of magnetized neutron star 
crust]{Effect of magnetized phonons on electrical and thermal 
conductivity of neutron star crust}

\author[D.A. Baiko]{D.A. Baiko\thanks{E-mail:baiko@astro.ioffe.ru} \\
A.F.\ Ioffe Physical-Technical Institute,
Politekhnicheskaya 26, 194021 St.-Petersburg, Russian Federation}

\begin{document}

\date{Accepted; Received ; in original form}

\pagerange{\pageref{firstpage}--\pageref{lastpage}} \pubyear{2013}

\maketitle

\label{firstpage}

\begin{abstract}
We study electrical and thermal conductivities of degenerate
electrons emitting and absorbing phonons in a strongly magnetized 
crystalline neutron star crust. We take into account  
modification of the phonon spectrum of a Coulomb solid of ions 
caused by a strong magnetic field. Boltzmann transport equation 
is solved using a generalized variational method. The ensuing 
three-dimensional integrals over the transferred momenta are evaluated 
by two different numerical techniques, the Monte-Carlo method and a 
regular integration over the first Brillouin zone. The results of 
the two numerical approaches are shown to be in a good agreement. 
An appreciable growth of electrical and thermal resistivities is 
reported at quantum and intermediate temperatures 
$T \lesssim 0.1 T_{\rm p}$ ($T_{\rm p}$ is the ion plasma temperature) 
in a wide range of chemical compositions and mass densities of matter 
even for moderately magnetized crystals 
$\omega_{\rm B} \sim \omega_{\rm p}$ ($\omega_{\rm B}$ and 
$\omega_{\rm p}$ are the ion cyclotron and plasma frequencies). This 
effect is due to an appearance of a soft ($\omega \propto k^2$) 
phonon mode in the magnetized ion Coulomb crystal, which turns out to 
be easier to excite than acoustic phonons characteristic of 
the field-free case. These results are important for modelling
magneto-thermal evolution of neutron stars.   
\end{abstract}

\begin{keywords}
dense matter -- stars: neutron.
\end{keywords}

\section{Introduction}
Magnificent seven is a group of isolated neutron stars emitting
quasi-thermal X-ray and optical radiation and situated at distances
below 500 pc. They have rotation periods in the 3--11 s range and 
magnetic fields and characteristic ages, estimated from the 
magneto-dipole braking formula, of (1--3)$ \times 10^{13}$ G and
(1--4)$ \times 10^6$ yrs, respectively. So, these are fairly old
nearby neutron stars, which, presumably, are seen solely due to their 
residual heat, but which all have magnetic fields higher than those 
of typical rotation-powered pulsars. Quite naturally, these stars were 
viewed as prime examples of objects, which are heated additionally 
by decay of their magnetic fields \citep[e.g.,][]{K10}. Further 
evidence in favor of this neutron star heating mechanism
was presented by \citet{PLMG07}, who have shown that there was a 
correlation between surface temperature and dipole magnetic 
field for such diverse objects as soft gamma repeaters, 
anomalous X-ray pulsars, isolated neutron stars, and rotation-powered
pulsars.

\citet{VRPPAM13} have performed detailed modelling of magneto-thermal 
evolution of neutron stars. In these simulations, magnetic 
field decay was able to provide enough heat to explain even the
hottest sources such as magnetars. Clearly, in order to obtain effective
decay of the magnetic field the electrical conductivity of matter must
be relatively low. To achieve that \citet{VRPPAM13} have assumed
that the nuclear pasta layer at the boundary of the inner crust with
the core possessed a high level of disorder. Accordingly, they
used impurity scattering with a very high impurity parameter
$Q_{\rm imp}=100$ ($Q_{\rm imp}=0.01$ in the rest of the crust) to 
model this situation.

Later on such a high value of the impurity parameter was justified in
a molecular dynamics simulation of the nuclear pasta layer by 
\citet{HBBCCS15}. For instance, in the lasagna phase, these authors have 
discovered a very peculiar defect, which looks like a spiral staircase
leading from one lasagna ``floor'' to the next and which may be 
responsible for $Q_{\rm imp} \approx 40$.

While this conjecture is certainly exciting, it brings about a number 
of questions. Firstly, one may wonder, how come there is so little 
disorder at lower and higher densities and so much disorder in the
narrow density range from $10^{14}$ to $1.5 \times 10^{14}$ g cm$^{-3}$.
Secondly, the high disorder layer then must be present also in
weakly magnetized neutron stars, where it is not required by 
observations. Thirdly, as shown by \citet{KYPSSG07}, it is more
economical to place a heat source of magnetars in more shallow layers
in order to avoid excessive energy loss via neutrinos, though these
considerations did not take into account proportional worsening
of the electron thermal conductivity in the high disorder layer.    
Ultimately, the question becomes whether it is possible to organize 
a fast field decay by more conventional means?    

In a crystalline neutron star crust charge and heat are transported by
degenerate electrons. The main mechanism impeding their transport is
emission and absorption of lattice 
vibrations (phonons) by electrons. Phonons in a magnetized
Coulomb crystal of ions (or ``magnetized phonons'') are fundamentally 
different from the field-free case \citep*[][]{UGU80,NF82,NF83,B09}.
In the absence of the field, the phonons are acoustic, i.e. their 
frequency depends linearly on the wave vector. In the presence of the
field, a soft mode with a quadratic dependence of the frequency on the
wave vector appears. 

The appearance of the soft mode can be understood
by referring to a simple problem of a charged oscillator with 
frequency $\omega_0$ (mass $m$ and charge $q$) brought into a uniform 
magnetic field $B$ \citep[e.g.,][]{LL94}. Its frequency becomes      
\begin{equation}
     \omega = \sqrt{\omega_0^2 + 
     \frac{1}{4} \omega_{\rm c}^2 } \pm
     \frac{1}{2} \omega_{\rm c}~,
\end{equation}
where $\omega_{\rm c} = qB/(mc)$.  
Assuming that $\omega_0 = c_{\rm s} k$, i.e. it represents an ion 
frequency in an acoustic phonon with sound speed $c_{\rm s}$,
the square root can be expanded at small wave vectors $k$, which 
results in the quadratic dependence 
$\omega \approx c^2_{\rm s} k^2 /\omega_{\rm c}$.
Clearly, the mode is softer for stronger magnetic fields.
This is precisely what one obtains if a lattice dynamics problem is 
solved for the Coulomb crystal in the magnetic field. 

The soft mode has a dramatic effect on the thermodynamic properties 
of the crystal at low temperatures. Due to its low frequency it is much 
easier to excite, which produces, for instance, an increase
of the crystal heat capacity per particle by a factor of $\sim 10^3$ at 
temperature $T=0.01 T_{\rm p}$ and phonon magnetization 
$b \equiv \omega_{\rm B}/\omega_{\rm p}=10$
\citep[cf.\ Fig.\ 2 of][]{B09}. In this case, 
$\omega_{\rm p} = \sqrt{4 \pi n Z^2 e^2/M}$ is the ion plasma 
frequency, $T_{\rm p} = \hbar \omega_{\rm p}/k_{\rm B}$ is the ion
plasma temperature (in what follows $k_{\rm B} = \hbar = c = 1$),
$\omega_{\rm B} = Z|e|B/(Mc)$ is the ion cyclotron frequency, while
$Z|e|$, $M$, and $n$ are ion charge, mass, and number density, 
respectively. It appears that the soft mode will
affect the effective rate of phonon 
emission/absorption by electrons in a comparable way. In fact, this 
rate [cf.\ Eq.\ (17) of \citet{BY95}, also see 
\citet{YU80,RY82}] is expressed 
via functions $G_0$ and $G_2$, which are given by similar averages over 
the phonon spectrum as the thermodynamic quantities and which are 
expected to grow strongly upon inclusion of the magnetic field. 
These simple considerations indicate that the problem of electron 
interaction with magnetized phonons in the Coulomb crystal deserves 
a serious study.

\section{Kinetic equation}
\label{KE}
Let us write the standard stationary Boltzmann kinetic equation
(see Sect.\ \ref{LOA} for a discussion of applicability limits):
\begin{eqnarray}
   {\bm v} \nabla f_{\bm p} + {\bm F} 
   \frac{\partial f_{\bm p}}{\partial {\bm p}} &=& 
   - \sum_{{\bm p}'{\bm k}} 
   \left\{\Gamma_{{\bm p}{\bm k}\to{\bm p}'} f_{\bm p} n_{\bm k} 
   (1-f_{{\bm p}'})
   +\Gamma_{{\bm p}\to{\bm p}'{\bm k}} f_{\bm p} (n_{\bm k}+1) 
   (1-f_{{\bm p}'}) \right.
\nonumber \\
   && ~~~~~ - \left. \Gamma_{{\bm p}'\to{\bm p}{\bm k}} f_{{\bm p}'} 
   (n_{\bm k}+1) (1-f_{\bm p})
    - \Gamma_{{\bm p}'{\bm k}\to{\bm p}} f_{{\bm p}'} n_{\bm k} 
    (1-f_{\bm p}) \right\}~.
\label{Boltz}
\end{eqnarray}
In this case, $f_{\bm p}$ is the electron momentum distribution 
function, which depends also on position ${\bm r}$ but is independent 
of the electron spin orientation,
${\bm v}$ is the electron velocity for momentum ${\bm p}$, and
${\bm F}$ is the Lorentz force: 
\begin{equation}
  {\bm F} = e{\bm E} +  e[{\bm v} \times {\bm B}]~,
\label{Lorentz}
\end{equation}
$e$, ${\bm E}$, and ${\bm B}$ being electron charge, electric and 
magnetic fields, respectively. On the right hand side of 
Eq.\ (\ref{Boltz}), 
$\Gamma_{{\bm p}{\bm k}\to{\bm p}'}$ is the transition probability
per unit time of electron with momentum ${\bm p}$ to a state
with momentum ${\bm p}'$ with an absorption of a phonon with
momentum ${\bm k}$ summed over primed and averaged over non-primed
electron spin states. Other $\Gamma$ refer to probabilities
of the other three possible processes of the same kind, while 
$n_{\bm k}$ is the phonon 
momentum distribution function. The summation is over all ${\bm p}'$, 
all ${\bm k}$ and over all phonon modes at given ${\bm k}$. For brevity, 
the phonon mode index $s$ is suppressed here.    

We linearize the Boltzmann equation assuming weak deviation of the
electron distribution from the local equilibrium
\begin{eqnarray}
     f_{\bm p} &=& f^0_{\bm p} + \delta f_{\bm p}~,
\nonumber \\ 
     f^0_{\bm p} &=& \left[\exp{\left(\frac{\varepsilon_{\bm p} - 
     \mu(r)}{T(r)}\right)} +1 \right]^{-1}~,
\label{df_def} 
\end{eqnarray}
and also assuming equilibrium phonon distribution 
\begin{equation}
     n_{\bm k} = n^0_{\bm k} = 
     \left[ \exp{\left(\frac{\omega_{\bm k}}{T(r)}\right)} 
     - 1 \right]^{-1}~,   
\end{equation}
where $\mu(r)$ and $T(r)$ are local electron chemical potential and
temperature, respectively, while $\varepsilon_{\bm p}$ and
$\omega_{\bm k}$ are electron energy and phonon frequency.  

For a strongly degenerate system it is customary \citep[e.g.,][]{Z60} 
to assume that the electron distribution deviates noticeably from the
local equilibrium one only near the Fermi surface:
\begin{equation}
     \delta f_{\bm p} = -\Phi_{\bm p} \frac{\partial f^0_{\bm p}}
     {\partial \varepsilon_{\bm p}} 
             = \frac{\Phi_{\bm p}}{T} \, f^0_{\bm p} (1-f^0_{\bm p})~,   
\end{equation}
where $\Phi_{\bm p}$ is a new unknown function. 

On the right-hand side of Eq.\ (\ref{Boltz}), the transition 
probabilities $\Gamma$ are determined by relativistic
electron scattering probability off a potential  
\begin{equation}
     \hat{U}({\bm r}) = \sum_I \int \frac{{\rm d}{\bm q}}{(2 \pi)^3} \, 
     \frac{4 \pi Z |e|}{q^2+\kappa_{\rm TF}^2}
     \, e^{i {\bm q} ({\bm r} - {\bm R}_I)} \, 
     (e^{-i {\bm q} \hat{\bm u}_I}-1)~.
\label{Upot}
\end{equation}
The potential is a sum of screened Coulomb potentials of all ions
(labeled by index $I$) minus the potential of the static lattice, in 
which all ions are fixed at their lattice nodes ${\bm R}_I$.
Furthermore, $Z$ is the ion 
charge number, $\kappa_{\rm TF}$ is the inverse Thomas-Fermi screening
length, and $\hat{\bm u}_I$ is the operator of ion displacement, which,
upon quantization of the ion motion, is given by \citep[][]{UGU80,B09}:    
\begin{equation}
    \hat{\bm u}_I = \frac{i}{\sqrt{MN}} \sum_{{\bm k}s} 
    (\bm{\alpha}_{{\bm k}s} \hat{a}_{{\bm k}s}
    - \bm{\alpha}_{{\bm k}s}^\ast \hat{a}_{-{\bm k}s}^\dagger)
    \, e^{i {\bm k}{\bm R}_I}~. 
\label{hatu}
\end{equation}
In this case, $N$ is the total number of ions, 
$\hat{a}^\dagger$ and $\hat{a}$ are phonon creation and 
annihilation operators, the sum is over all phonon modes, while
vectors $\bm{\alpha}_{{\bm k}s}= \bm{\alpha}_{-{\bm k}s}$ are 
analogous to phonon polarization vectors in a non-magnetized crystal, 
but have different orthogonality and normalization properties 
\cite[see][for details]{B09}.  

We restrict ourselves to the one-phonon approximation, in which
only the first order term in $\hat{\bm u}_I$ is kept in 
Eq.\ (\ref{Upot}). Multi-phonon processes become important closer to 
the crystal melting temperature \citep[][]{BKPY98}, 
whereas we are mostly focused
on lower temperatures. Then various $\Gamma$ on the right-hand 
side of Eq.\ (\ref{Boltz}) differ only by the energy-conserving 
delta-functions and the linearized kinetic equation reads:
\begin{eqnarray}
      &-& \frac{\partial f^0_{\bm p}}{\partial \varepsilon_{\bm p}} 
      \left( \frac{\varepsilon_{\bm p} - \mu}{T} {\bm v}{\nabla T}
      + {\bm v}{\nabla \mu} - {\bm v}(e {\bm E}) 
      + e[{\bm v} \times {\bm B}]
     \frac{\partial \Phi_{\bm p}}{\partial {\bm p}} \right)     
\nonumber \\
      &=& - \sum_s \int {\rm d} {\bm p}' \frac{4nZ^2 e^4}
      {M (q^2 + \kappa_{\rm TF}^2)^2} |{\bm q}\bm{\alpha}_{{\bm k}s}|^2
      \left(1- \frac{v_{\rm F}^2 q^2}{4 p_{\rm F}^2} \right) \times 
\nonumber \\      
      &\times& f_{\bm p}^0 (1-f_{{\bm p}'}^0) 
      [n_{{\bm k}s}^0 \delta(\varepsilon_{\bm p} + \omega_{{\bm k}s} 
      - \varepsilon_{{\bm p}'})
      + (n_{{\bm k}s}^0+1) \delta(\varepsilon_{\bm p} 
      - \omega_{{\bm k}s} - \varepsilon_{{\bm p}'})]
      \frac{(\Phi_{\bm p}- \Phi_{{\bm p}'})}{T}~.
\label{linKE}
\end{eqnarray}
In this case, $v_{\rm F}$ and $p_{\rm F}$ are electron Fermi velocity 
and momentum, ${\bm q} = {\bm p}' - {\bm p}$, and 
${\bm k} = {\bm q} - {\bm G}$, where ${\bm G}$ is a reciprocal lattice 
vector chosen in such a way that ${\bm k}$ is in the first Brillouin 
zone. Let us note, 
that the electron distribution responds to $\nabla \mu$ and to the 
electric field ${\bm E}$ in exactly the same way, which allows us 
to set $\nabla \mu=0$ without any loss of generality.

\section{Solution by variational principle}
\label{VP}
Typically, in a magnetic field, one adopts the relaxation time 
approximation for the collision integral on the right-hand side
of Eq.\ (\ref{Boltz}) or (\ref{linKE}), where the relaxation 
time is taken from
the respective non-magnetic problem \citep[e.g.,][]{Z60,A70,UY80}. 
This approach is well justified
if we do not expect the collision probability to depend on $B$ and 
also the collisions are nearly elastic (i.e. the energy difference 
between the initial and the final electron states is much lower than 
$T$). In our case, both of these assumptions are invalid, since the
scatterer (i.e. phonons') properties depend on $B$, and the most 
interesting effect is anticipated at low temperatures, where the details 
of the phonon spectrum are important and the transferred energy is 
of the order of $T$. 

In the field-free case the variational principle has been used 
successfully \citep[e.g.,][]{Z60,FI76,RY82} to study the transport 
properties at low 
temperatures. There also exists a generalization of the variational 
principle to the case of non-zero magnetic field \citep[][]{Z60}. 
The kinetic 
equation (\ref{linKE}) can be written in a symbolic form as  
\begin{equation}
      X = P \Phi + M({\bm B}) \Phi~,
\label{symbKE}
\end{equation}
where $\Phi$ is the unknown function, $X$ denotes the terms
on the left-hand side which drive the system out of the equilibrium 
(i.e. terms with an electric field or a temperature gradient),
$P$ is the collision operator, and $M({\bm B})$ is the magnetic
operator, which contains the momentum derivative of $\Phi$ and
which is moved from the left-hand side to the right. 

The variational solution is looked for in the form of a linear 
combination of some basis functions $\phi_i({\bm p})$: 
\begin{equation}
        \Phi = \sum_i \tau_i \phi_i({\bm p})~,
\label{ansatz_gen}
\end{equation}
where $\tau_i$ are unknown constants. These constants are solution
of a system of linear equations
\begin{equation}
       \langle \phi_i , X\rangle = 
       \sum_j \langle \phi_i , P \phi_j\rangle \tau_j
       + \sum_j \langle \phi_i , M({\bm B}) \phi_j\rangle \tau_j~,  
\label{system}
\end{equation}
in which angle brackets denote a scalar product 
\begin{equation}
      \langle \phi, \psi \rangle \equiv 
      \int {\rm d} {\bm p} \, \phi({\bm p}) \psi({\bm p})~.
\label{scal_prod}
\end{equation}

In our problem the form of the unknown function 
Eq.\ (\ref{ansatz_gen})
is suggested by the solution of the kinetic equation in the
relaxation time approximation \citep[e.g.,][]{A70}, i.e.
\begin{equation}
   \Phi_{\bm p} = \tau_1 {\bm v}\cdot(e{\bm E}) 
   + \tau_2 {\bm v}\cdot \frac{{\bm B} (e{\bm E}\cdot {\bm B})}{B^2}
   + \tau_3 {\bm v} \cdot \frac{[e{\bm E} \times {\bm B}]}{B}
\label{a_sig}
\end{equation}
for the charge transport problem and
\begin{equation}
   \Phi_{\bm p} = -\frac{(\varepsilon_{\bm p} - \mu)}{T} \left\{
   \tau_1 {\bm v}\cdot \nabla T + \tau_2 {\bm v}\cdot \frac{{\bm B} 
   (\nabla T \cdot {\bm B})}{B^2}
   + \tau_3 {\bm v} \cdot \frac{[\nabla T \times {\bm B}]}{B} \right\}
\label{a_kap}
\end{equation}
for the heat transport problem. Then the scalar products involving $X$
and $M({\bm B})$ become trivial, for instance
\begin{eqnarray}
     \langle \phi_i , M({\bm B}) \phi_j \rangle &=& 
     \int {\rm d}{\bm p} \,
     ({\bm v}\cdot{\bm u}_i) \left(-\frac{\partial f^0_{\bm p}}
     {\partial \varepsilon_{\bm p}} \right) 
      [{\bm v} \times {\bm B}] \cdot \frac{\partial}{\partial {\bm p}} 
      ({\bm v}\cdot{\bm u}_j)
\nonumber \\
      &=&  \frac{4\pi}{3} {\bm u}_j\cdot[{\bm u}_i \times {\bm B}]
      \int {\rm d} p \, \frac{p^2 v^2}{\varepsilon_{\bm p}} 
      \left(-\frac{\partial f^0_{\bm p}}{\partial \varepsilon_{\bm p}} 
      \right)  =  \frac{4\pi}{3} \, p_{\rm F} v_{\rm F}^2 \, 
      {\bm u}_j\cdot[{\bm u}_i \times {\bm B}]~, 
\end{eqnarray}
where ${\bm u}_i$ denote various constant vectors, 
which appear in the scalar products with the velocity in 
Eqs.\ (\ref{a_sig}) and (\ref{a_kap}). 

By contrast, the scalar products involving the collision operator $P$
are not easy to evaluate. They contain 6D-integrals over ${\bm p}$
and ${\bm p}'$, of which the integrals over $p$ and $p'$ can be 
taken by the standard methods. The remaining 4D-integrals over the 
solid angles $\Omega_{\bm p}$ and $\Omega_{{\bm p}'}$ can be reduced 
to 3D-integrals over the transferred momentum ${\bm q}$ over a ball of
radius $2p_{\rm F}$ as follows        
\begin{eqnarray}
   \int {\rm d}\Omega_{\bm p} {\rm d}\Omega_{{\bm p}'} &=& 
   \int {\rm d}\Omega_{\bm p} {\rm d}\Omega_{{\bm p}'} 
   \int {\rm d} p' \, \frac{p^{\prime 2}}{p^2_{\rm F}} \, 
   \delta(p'-p_{\rm F})  
   \int_{{\rm ball}<2p_{\rm F}} {\rm d} {\bm q} \, 
   \delta({\bm q}+{\bm p}-{\bm p}')
\nonumber \\
  &=& \int_{\rm ball} {\rm d} {\bm q} \, \int {\rm d}\Omega_{\bm p} \,
  \frac{1}{p^2_{\rm F}} \,
  \delta(|{\bm q}+{\bm p}|-p_{\rm F}) = \frac{2 \pi}{p^2_{\rm F}}
  \int_{\rm ball} \frac{{\rm d} {\bm q}}{q}~.    
\label{3D-int}
\end{eqnarray}
In the process, we have integrated over the azimuthal angle of vectors 
${\bm p}$ and ${\bm p}'$ with respect to ${\bm q}$. The presence
of this integration allows one to replace without any loss of
accuracy
\begin{equation}
  \int {\rm d}\Omega_{\bm p} {\rm d}\Omega_{{\bm p}'} 
  v_{\alpha}  (v_\beta - v'_\beta) \, u_{i\alpha} u_{j\beta} \to   
   \int {\rm d}\Omega_{\bm p} {\rm d}\Omega_{{\bm p}'} 
   \frac{q_\alpha q_\beta}{2 \varepsilon_{\rm F}^2}  \, 
   u_{i\alpha} u_{j\beta}~, 
\end{equation}
for the electrical conductivity, and make similar replacements for
the thermal conductivity ($\varepsilon_{\rm F}$ is the electron Fermi 
energy).  

Combining everything together we obtain for the electrical and 
thermal conductivities ($\sigma$ and $\kappa$) the following systems 
of linear equations, respectively:
\begin{eqnarray}
      e{\bm E}\cdot{\bm u}_i &=& \sum_j \tau_j 
      \left(\nu^\sigma_{\alpha\beta} 
      u_{i\alpha} u_{j\beta} + \omega_{\rm eB} {\bm b} \cdot 
      [{\bm u}_i \times {\bm u}_j ] \right)~,
\nonumber \\
      \nabla T \cdot{\bm u}_i &=& \sum_j \tau_j 
      \left(\nu^\kappa_{\alpha\beta} 
      u_{i\alpha} u_{j\beta} + \omega_{\rm eB} {\bm b} \cdot 
      [{\bm u}_i \times {\bm u}_j ] \right)~,
\label{sle}
\end{eqnarray}
with ${\bm b} = {\bm B}/B$ and $\omega_{\rm eB}$ being the electron 
gyro-frequency  
\begin{equation}
   \omega_{\rm eB} = \frac{|e|B}{\varepsilon_{\rm F}}~. 
\end{equation}
The quantities $\nu^{\sigma,\kappa}_{\alpha\beta}$ can be called 
collision frequency tensors. They read
\begin{equation}
       \nu^\sigma_{\alpha\beta} = \frac{3e^2}{\hbar v_{\rm F}} \, 
       \frac{2T}{(4\pi)^2 t^2} \sum_s \int {\rm d}\Omega_{\bm p} 
       {\rm d}\Omega_{{\bm p}'} \,
       \frac{2 \omega_{{\bm k}s} |{\bm q}\bm{\alpha}_{{\bm k}s}|^2}
       {(q^2 + \kappa_{\rm TF}^2)^2} 
      \left(1- \frac{v_{\rm F}^2 q^2}{4 p_{\rm F}^2} \right) 
      \frac{e^{\omega_{{\bm k}s}/T}}{(e^{\omega_{{\bm k}s}/T}-1)^2} \, 
      q_{\alpha} q_{\beta}~,
\label{nusig_ab}
\end{equation}
and $\nu^\kappa_{\alpha\beta}$ differs by the replacement
\begin{equation}
    q_{\alpha} q_{\beta} \to q_{\alpha} q_{\beta} 
    \left( 1+ \frac{\omega^2_{{\bm k}s}}{4 \pi^2 T^2} \right)
    + \frac{\omega^2_{{\bm k}s}}{T^2} 
    \left(\delta_{\alpha\beta} -\frac{q_\alpha q_\beta}{q^2} \right)
      \frac{3 p^2_{\rm F} }{2 \pi^2}
    \left(1-\frac{q^2}{4p^2_{\rm F}} \right)~.
\end{equation}
In principle, Eqs.\ (\ref{sle}) can be solved easily for arbitrary 
orientations of the magnetic field, electric field, and temperature 
gradient with respect to the crystal axes. However, we shall assume that
the crystal always forms in such a way that the magnetic field is 
directed along the symmetry axes resulting in the minimum 
zero-point energy \citep[for bcc crystal this would be a direction 
towards
one of the nearest neighbors,][]{B09}. At the same time, we would like 
to 
average over the azimuthal angle of the electric field or temperature
gradient with respect to the magnetic field. This can be done in
various ways. For instance, one can average the effective collision 
frequency or the effective relaxation time, or the kinetic coefficients
themselves. It is not immediately clear, which kind of averaging should 
be preferred. We note, that the averaging procedure may affect the 
final practical results in a non-trivial manner \citep[cf.\ ][]{KP15}. 

In this paper we adopt the simplest approach and average the system of 
linear equations (\ref{sle}) itself. Then for the electrical 
conductivity we obtain (upper index $\sigma$ is understood for all 
$\tau$ and $\nu$)  
\begin{equation}
    \tau_1 = \frac{\tau_\perp}
              {1+\omega_{\rm eB}^2 \tau^2_\perp} ~~~~~~
    \tau_2 = \tau_\parallel - \tau_1 ~~~~~~
    \tau_3 = \frac{\omega_{\rm eB} \tau^2_\perp}
    {1+\omega_{\rm eB}^2 \tau^2_\perp}~, 
\end{equation}
where
\begin{equation}
    \tau_\perp = \frac{1}{\nu_\perp}  ~~~~~  
    {\nu_\perp} = \frac{1}{2} \nu_{\alpha\beta} 
    (\delta_{\alpha\beta}-b_\alpha b_\beta) ~~~~~
    \tau_\parallel = \frac{1}{\nu_\parallel}  ~~~~~ 
    \nu_\parallel = \nu_{\alpha\beta} b_\alpha b_\beta~. 
\label{perp_paral}
\end{equation}

Once the deviation of the distribution function from the local
equilibrium is known
\begin{equation}
   \Phi_{\bm p} = e (\tau_1+\tau_2) \, 
   ({\bm v} \cdot {\bm E}_{\parallel})
     + e \tau_1 ({\bm v} \cdot {\bm E}_{\perp})
     + e \tau_3 ({\bm v} \cdot [{\bm E}_{\perp} \times {\bm b}])~,
\end{equation}
where ${\bm E}_{\parallel}$ and ${\bm E}_{\perp}$ refer to components 
of the electric field parallel and perpendicular 
to the magnetic field, one can calculate the electric current
\begin{equation}
    J_\alpha = \sigma_{\alpha \beta} E_\beta = 2 \int 
    \frac{{\rm d}{\bm p}}{(2\pi)^3} \, \delta f_{\bm p} \, e v_\alpha
    = 2 \int \frac{{\rm d}{\bm p}}{(2\pi)^3} \, \Phi_{\bm p}
\left(-\frac{\partial f^0_{\bm p}}{\partial \varepsilon_{\bm p}} \right)
    e v_\alpha
\end{equation}
and determine the components of the electrical conductivity tensor
\begin{equation}
   \sigma_{xx}=\sigma_{yy} = \frac{e^2 n_e}{\varepsilon_{\rm F}/c^2}\,
   \frac{\tau_\perp}{1+\omega_{eB}^2 \tau_\perp^2}~, ~~~ 
   \sigma_{zz}= \frac{e^2 n_e}{\varepsilon_{\rm F}/c^2} \,
   \tau_\parallel~, ~~~
   \sigma_{xy}=-\sigma_{yx} = \frac{e^2 n_e}{\varepsilon_{\rm F}/c^2}\,
    \frac{\omega_{eB} \tau_\perp^2}{1+\omega_{eB}^2 \tau_\perp^2}~.   
\label{sigmas}
\end{equation}
In this case it is assumed that the $z$-axis is directed along 
${\bm B}$. Thus
the perpendicular time $\tau_\perp$ determines the conductivity 
across the magnetic
field, the parallel time $\tau_\parallel$ determines the conductivity 
along the magnetic 
field, and $\tau_3$ determines the Hall conductivity. In the relaxation
time approximation one obtains exactly the same formulae with
$\tau_\perp$ and $\tau_\parallel$ replaced by the nonmagnetic
relaxation time $\tau_0$ \citep[][]{UY80}.

For the heat flux we get:
\begin{equation}
    Q_\alpha = - \kappa_{\alpha \beta} \nabla_\beta T = 2 \int 
    \frac{{\rm d}{\bm p}}{(2\pi)^3} \, \delta f_{\bm p} \, 
    (\varepsilon_{\bm p} - \mu) v_\alpha =
   2 \int  \frac{{\rm d}{\bm p}}{(2\pi)^3} \, \Phi_{\bm p}
\left(-\frac{\partial f^0_{\bm p}}{\partial \varepsilon_{\bm p}} \right) 
    (\varepsilon_{\bm p} - \mu) v_\alpha~,
\end{equation}
and the components of the thermal conductivity tensor read
\begin{equation}
   \kappa_{xx}=\kappa_{yy} = \frac{\pi^2 n_e T}
   {3 \varepsilon_{\rm F}/c^2}\,
   \frac{\tau_\perp}{1+\omega_{eB}^2 \tau_\perp^2}~, ~~ 
   \kappa_{zz}= \frac{\pi^2 n_e T}{3\varepsilon_{\rm F}/c^2} \,
   \tau_\parallel~, ~~
   \kappa_{xy}=-\kappa_{yx} = \frac{\pi^2 n_e T}
   {3 \varepsilon_{\rm F}/c^2}\,
    \frac{\omega_{eB} \tau_\perp^2}{1+\omega_{eB}^2 \tau_\perp^2}~,   
\label{kappas}
\end{equation}
where it is understood that $\tau^\kappa$ must be used.  
To obtain $\tau^\kappa$ from $\tau^\sigma$ the following replacements 
have to be made
in the expressions (\ref{perp_paral}) for parallel and perpendicular 
frequencies $\nu_{\parallel,\perp}$ 
\begin{eqnarray}
   q_{\alpha} q_{\beta} b_\alpha b_\beta \equiv
   q_\parallel^2 &\to& q_\parallel^2 
   \left(1+ \frac{\omega^2_{{\bm k}s}}{4\pi^2T^2}\right)
     + \frac{\omega^2_{{\bm k}s}}{T^2} 
     \left(1-\frac{q_\parallel^2}{q^2} \right)  
     \frac{3 p^2_{\rm F}}{2 \pi^2} 
     \left(1-\frac{q^2}{4p_{\rm F}^2}\right)~, 
\nonumber\\
   q^2 - q_\parallel^2 \equiv q_\perp^2 &\to& 
   q_\perp^2 \left(1+ \frac{\omega^2_{{\bm k}s}}
   {4\pi^2T^2}\right)
     + \frac{\omega^2_{{\bm k}s}}{T^2} \left(1+
     \frac{q_\parallel^2}{q^2} \right)  
     \frac{3 p^2_{\rm F}}{2 \pi^2} 
     \left(1-\frac{q^2}{4p_{\rm F}^2}\right)~. 
\end{eqnarray}

\section{Numerical calculations}
\begin{figure}
\begin{center}
\leavevmode
\includegraphics[bb=13 13 613 550, width=84mm]{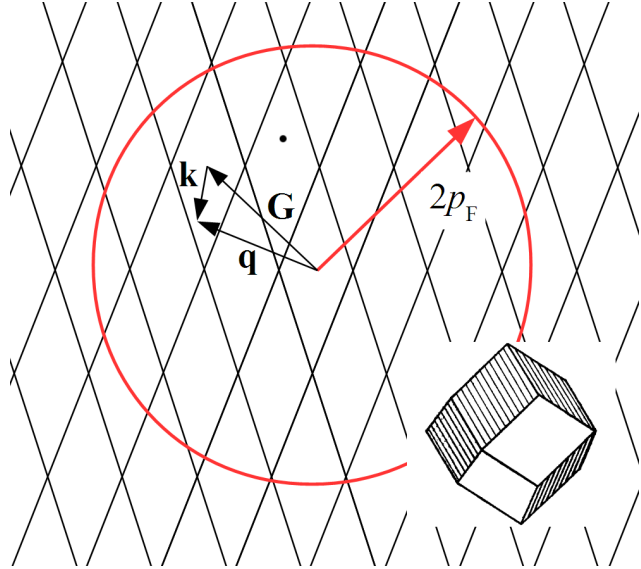}
\end{center}
\vspace{-0.4cm}
\caption[ ]{Two-dimensional sketch of the reciprocal space. The  
circle and rhombi represent the double Fermi sphere and the first 
Brillouin zone shifted by all possible reciprocal lattice vectors, 
respectively. The actual 3D first Brillouin zone is shown in the 
inset. For any given transferred momentum ${\bm q}$, the phonon wave 
vector ${\bm k} = {\bm q} - {\bm G}$, where ${\bm G}$ is the 
appropriate reciprocal lattice vector.}
\label{ball}
\end{figure}

The most difficult part of the problem is to evaluate 3D-integrals
Eq.\ (\ref{3D-int}) over the transferred momentum ${\bm q}$ over
the ball of radius $2p_{\rm F}$. In Fig.\ \ref{ball} a sphere of 
radius $2p_{\rm F}$ (``double Fermi sphere'') is shown by the (red) 
circle, while rhombi represent the first Brillouin zone shifted 
by all possible reciprocal lattice vectors, which results in 
a complete filling of the momentum space. 
The actual (three-dimensional) first Brillouin zone of the bcc lattice 
is depicted in the inset (a rhombododecahedron). A reciprocal lattice 
vector connects the center of the sphere with the center of each 
rhombus. In order to find the phonon frequencies and the 
$\bm{\alpha}$-coefficients for a given ${\bm q}$, we need to subtract 
the respective reciprocal lattice vector ${\bm G}$ to obtain a phonon
wave vector ${\bm k} = {\bm q} - {\bm G}$ in the first Brillouin zone,
and then solve a lattice dynamics problem at this ${\bm k}$.

Thus, essentially, we need to integrate over the first Brillouin zone.
Methods of such integration in a Coulomb solid are well-developed
\citep*[e.g., Holas method,][]{H77,AG81,BPY01},
however, this is only true if we need to integrate over the whole
Brillouin zone. Unfortunately, as is evident from Fig.\ \ref{ball}, 
there are numerous incomplete pieces of the Brillouin zones inside
the ball. They appear due to the intersections of the double Fermi 
sphere with the zones. It seems that the problem of enumerating all 
these pieces and integrating over them is too cumbersome to deal with. 
Thus, the first approach that comes to mind 
\citep[see also][]{RY82,BY95} is to use the Monte-Carlo (MC) 
integration method, i.e. to find the average of the integrand over 
the ball by randomly selecting points inside the ball.

\begin{figure}
\begin{center}
\leavevmode
\includegraphics[bb=7 17 350 348, width=84mm]{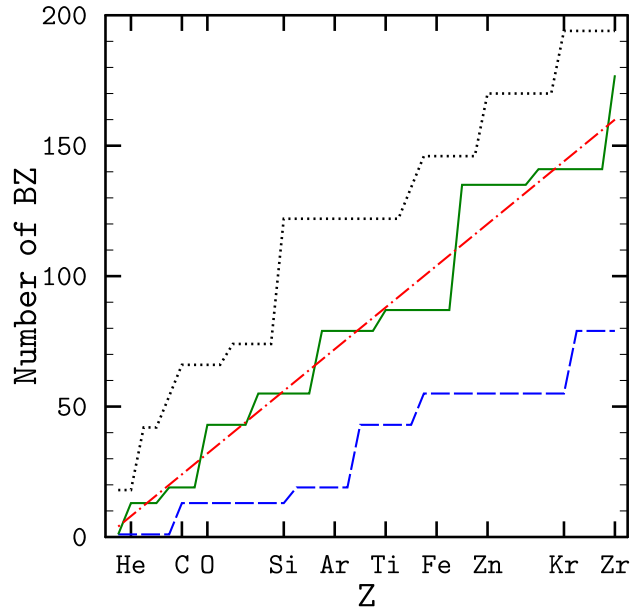}
\end{center}
\vspace{-0.4cm}
\caption[ ]{Number of Brillouin zones intersecting with the double 
Fermi sphere (dotted, black), having their centers inside the sphere 
(solid, green), and lying fully inside the sphere (dashed, blue)
as functions of the atomic charge number $Z$. Straight dash-dotted 
(red) line $y=4Z$ is the volume of the sphere in units of the 
Brillouin zone volume.}
\label{BZs}
\end{figure}

At low temperatures, though, the integrand develops very sharp
and narrow maxima in the vicinity of points ${\bm q}={\bm G}$. This 
means that one needs many more MC coin tosses to get an adequate 
representation of the integrand (in order to reduce temperature 10 
times, $\sim 1000$ times more MC steps is required). Another approach 
to this integration, which is suited better to lower temperatures, 
is thus desired and it is illustrated in Fig.\ \ref{BZs}. 
In this picture, the dash-dotted (red) line $y=4Z$ is the volume 
of the double Fermi ball measured in volumes of the Brillouin zone. 
The dashed (blue) line is the number of Brillouin zones, which lie 
wholly inside the ball, and the dotted (black) line is the number 
of Brillouin zones, which intersect the ball (i.e. it is the number 
of incomplete zones inside the ball). Finally, the solid (green) line
is the number of Brillouin zones, which have their centers inside 
the ball. It is clear that the total volume of the latter zones is 
a reasonable approximation of the ball volume. We may thus try to 
integrate over these zones, taking them wholly, and expect this to be
a very accurate result at low temperatures as all points 
${\bm q}={\bm G}$ will be accounted for. At higher temperatures, 
the MC method must be more precise because of the proper 
integration domain used in this case.

\begin{figure}
\begin{center}
\leavevmode
\includegraphics[bb=9 3 348 344, width=84mm]{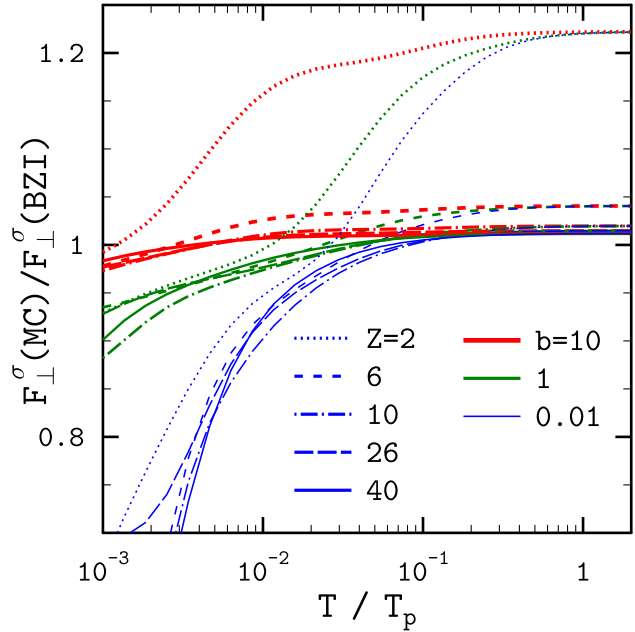}
\end{center}
\vspace{-0.4cm}
\caption[ ]{The ratio of $F^{\sigma}_{\perp}$ calculated by MC 
and BZI methods as a function of $T/T_{\rm p}$ for ultrarelativistic 
electrons and for $b=0.01$ (thin blue lines), 1 (green lines of 
intermediate thickness), and 10 (thick red lines). Dotted, short-dashed, 
dash-dotted, long-dashed, and solid lines correspond to $Z=2$, 6, 10, 
26, and 40, respectively.}
\label{int-methods}
\end{figure}

Let us compare integration results by the two methods.
We define dimensionless functions 
$F^{\sigma,\kappa}_{\parallel,\perp} \equiv 
\nu^{\sigma,\kappa}_{\parallel,\perp} \hbar^2 v_{\rm F}/(T e^2)$
\citep[cf.][]{BY95} and plot in Fig.\ \ref{int-methods} the ratio
of $F^\sigma_\perp$ calculated by the MC method to the same quantity
calculated via the Brillouin zone integration (BZI). The ratios are 
plotted as functions of $T/T_{\rm p}$ for several
ion charge numbers $Z=2$ (dotted), 6 (short-dashed), 10 (dash-dotted), 
26 (long-dashed), 40 (solid) and phonon magnetizations 
$b=\omega_{\rm B}/\omega_{\rm p}=0.01$ (thin, blue), 
1 (intermediate thickness, green), 10 (thick, red). 
In all cases electrons are assumed to be ultrarelativistic. 
For $Z=6$ and 10 the data are based on $10^6$ MC steps, while 
for $Z=2$, 26, and 40 this number is doubled.

We note that the ratios look rather insensitive to the charge number
with the exception of helium (dots). The lowest 
magnetization $b=0.01$ curves display a sharp relative decrease 
of MC integrals at $T \lesssim 0.01 T_{\rm p}$. In this temperature 
range BZI results are robust. They are insensitive to the number
of integration points in the Holas method, provided it is not too 
small. By contrast, MC results at these low temperatures are very 
sensitive to the number of MC steps due to the effect described above. 
At higher magnetizations though, the MC results converge much better, 
because wider regions around the Brillouin zone center 
${\bm q}={\bm G}$ continue to contribute. 
At $T \gtrsim 0.01 T_{\rm p}$, MC calculations, presumably, are 
more reliable. The deviation of the BZI values can be explained by 
different geometry of the overall integration domains (ball vs. a set 
of whole Brillouin zones). At some intermediate temperatures MC and BZI 
coincide and the ratio becomes 1. In principle, one can use BZI 
below this point and MC above it. However, the inaccuracy of the 
BZI results at high temperatures is so insignificant ($<5\%$ for all 
elements except helium) that we have decided to drop
MC calculations altogether and use BZI in the entire 
temperature range. Only for low-$Z$ elements such as helium the 
accuracy of BZI at $T \gtrsim 0.01 T_{\rm p}$ becomes
somewhat insufficient ($\lesssim 20\%$) and, in principle, the 
MC approach should be preferred.

The same graphs could be constructed for $F^\sigma_\parallel$ as well 
as for $F^\kappa_{\parallel,\perp}$ but they would reveal nothing 
principally new.
   
\begin{figure*}
\begin{center}
\leavevmode
\includegraphics[bb=72 410 569 741, width=170mm]{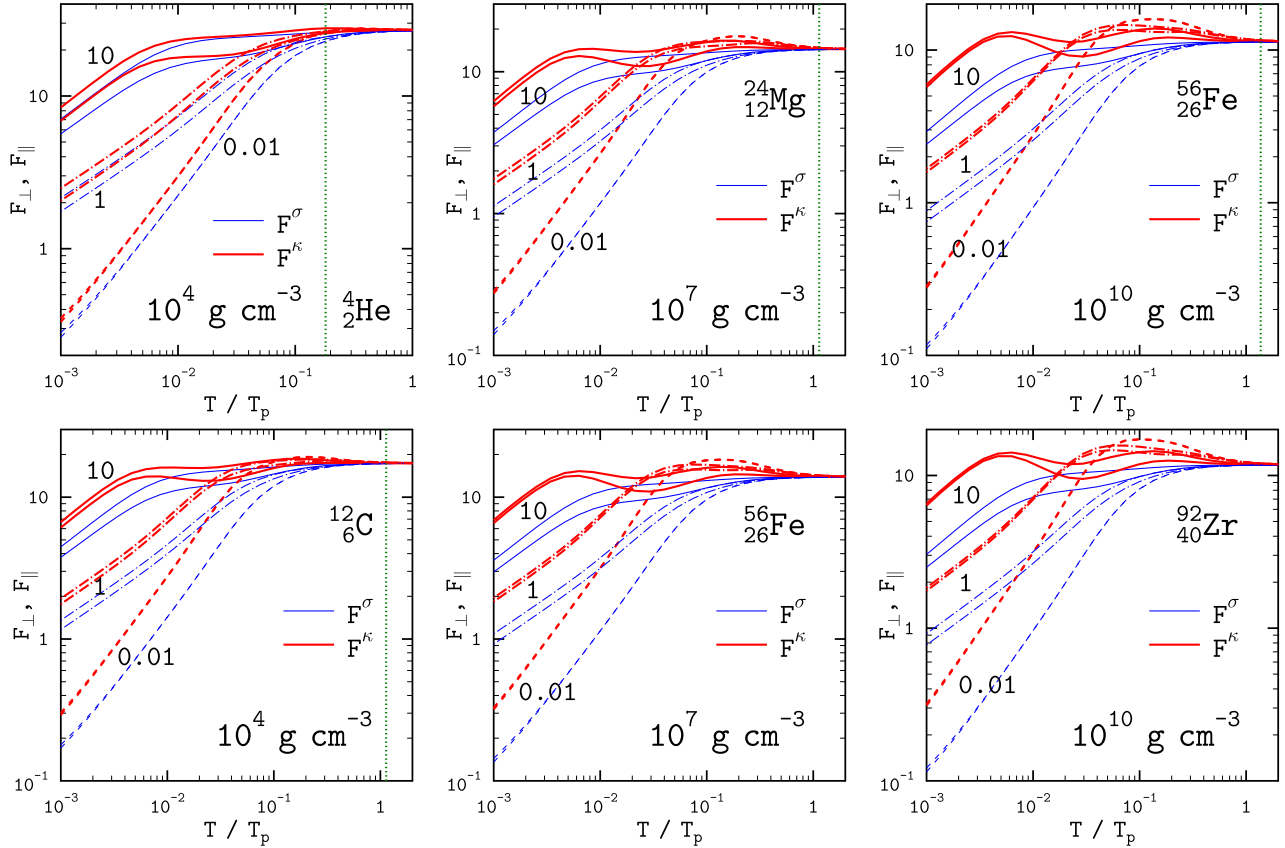}
\end{center}
\vspace{-0.4cm}
\caption[ ]{$F^{\sigma,\kappa}_{\parallel,\perp}$ calculated 
using BZI versus $T/T_{\rm p}$
for several chemical elements, mass densities, and phonon 
magnetizations. Thin (blue) curves illustrate
$F^\sigma$, while thick (red) ones show $F^\kappa$. Dashed, dash-dotted, 
and solid lines correspond to $b=0.01$, 1, and 10, respectively. 
For each value of $b$ two lines are shown for both $F^\sigma$ and 
$F^\kappa$: the one with larger values corresponds to $F_\perp$ and 
the other one to $F_\parallel$ (notice that for
$b=0.01$ the two lines practically coincide). Dotted
(green) lines show melting temperatures.}
\label{results}
\end{figure*}

Our final results are presented in Fig.\ \ref{results}. These plots 
show quantities $F^{\sigma,\kappa}_{\parallel,\perp}$ (directly 
proportional to the collision frequencies) as functions of
$T/T_{\rm p}$ for several representative of the outer neutron star
crust chemical elements ranging from $^{4}_{2}$He to $^{92}_{40}$Zr 
and mass densities $\rho$ ranging from 
$10^4$ to $10^{10}$ g cm$^{-3}$. Thin (blue) and thick (red) curves 
show $F^\sigma$ and $F^\kappa$, respectively. Vertical dotted lines 
show the melting temperature if it belongs to the displayed range of 
temperatures (calculations are artificially extended beyond melting 
for illustrative purposes).  
The ion magnetization parameter $b$ is set to 0.01, 1, and 10 for 
dashed, dash-dotted, and solid lines, respectively, and is
marked near the curves.  
We note that at given $T/T_{\rm p}$, $b$, and composition, the 
mass density determines the electron degree of relativity and enters 
Eq.\ (\ref{nusig_ab}) only through the Thomas-Fermi screening length 
and the back-scattering suppression factor in big parentheses. 

The lowest magnetization $b=0.01$ curves essentially reproduce earlier 
field-free calculations (except at extremely low temperatures 
$T\lesssim 10^{-3} T_{\rm p}$, where even such a weak magnetic field 
starts making a difference). Accordingly, for $b=0.01$ parallel and 
perpendicular $F$-functions merge. At high temperatures 
$T \gtrsim T_{\rm p}$, all curves merge, which means that the 
scattering is quasi-elastic and the exact phonon spectrum is not 
important. At intermediate and low 
temperatures and $b \gtrsim 1$ one observes a divergence of parallel 
and perpendicular frequencies ($F_\perp > F_\parallel$ in all 
cases) as well as their significant growth as 
compared to the low magnetization case. This produces a 
proportional decrease of the electrical and thermal conductivities and 
represents the main result of our work. 

The increase of the collision frequencies is not as strong as one
would expect from the comparison with the specific heat in 
the Introduction. This can be explained by a suppression of 
the amplitude Eq.\ (\ref{hatu}) of ion deviation from its lattice 
node in a strong magnetic field and a corresponding reduction of 
the phonon potential Eq.\ (\ref{Upot}), whereas the argument in the 
Introduction took into account only statistical weight of phonons.

\section{Limits of applicability}
\label{LOA}
While solving the transport problem, we have introduced 
several important simplifications. First of all, we have defined 
momentum distribution function for electrons $f_{\bm p}$, 
Eq.\ (\ref{Boltz}), as opposed to considering distribution of electrons 
over quantum numbers appropriate in the magnetic field \citep[i.e., 
longitudinal momentum, Landau level number etc., see, e.g.,][]{Y84}. 
Secondly, 
we have neglected a magnetic field effect on the electron 
screening of the ion potential, Eq.\ (\ref{Upot}). Thirdly, we have 
neglected an effect (not studied to this day) of electron screening
(with or without field) on the magnetized Coulomb crystal phonons. 

\begin{figure}
\begin{center}
\leavevmode
\includegraphics[bb=6 11 360 348, width=84mm]{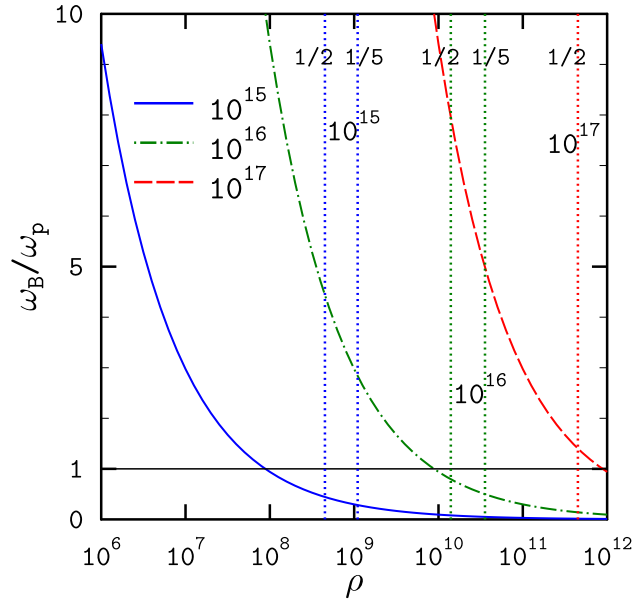}
\end{center}
\vspace{-0.4cm}
\caption[ ]{Phonon magnetization parameter 
$b=\omega_{\rm B}/\omega_{\rm p}$ as a function of mass 
density $\rho$ for $B=10^{15}$ (solid, blue), $10^{16}$ (dash-dotted, 
green), and $10^{17}$ (dashed, red) Gauss. Dotted vertical lines
mark the densities at which the ground Landau level is completely
filled for $Z/A=1/2$, $Z/A=1/5$ and the same magnetic fields.}
\label{limits}
\end{figure}

It is well-known that actual kinetic coefficients in a magnetic field
oscillate, e.g. as functions of density at given temperature 
\citep[e.g.,][]{Y84,P96,P99}. The lower is the temperature, the stronger 
are the oscillations. However, if more than one Landau level 
is populated,
one may expect that our consideration based on the momentum distribution
$f_{\bm p}$ yields an average over the oscillations value of the 
kinetic coefficients. The actual coefficients will trace this average
more closely as temperature gets higher. The same argument allows us 
to neglect the effect of electron screening modification by the 
magnetic field for electrons populating more than one Landau level. 
The effect of electron screening on magnetized 
phonons has not been analysed yet. However, in the absence of the field 
the Coulomb crystal phonons are modified by electron screening in a 
well-studied way \cite[][]{PH73,B02} and this makes a barely visible 
imprint on kinetics \cite[][]{BY95}. 
While caution is required, we can expect the same conclusion to hold 
in the presence of the magnetic field as well.   
 
If electrons populate only the ground Landau level, our approach becomes 
unreliable, and the problem of electron transport must be reconsidered
with account of magnetized phonons, inelasticity of scattering, and 
electron screening. This problem is further complicated by the fact
that the Fermi temperature of electrons populating only the ground 
Landau level drops very rapidly with decrease of density. Consequently, 
such electrons become non-degenerate at relevant 
temperatures and densities. Especially problematic also is a quick 
onset of the strong screening regime, which accompanies the drop of 
the Fermi temperature and electron kinetic energy.  

In Fig.\ \ref{limits} we show the phonon magnetization parameter $b$
as a function of density for $B=10^{15}$, $10^{16}$, and $10^{17}$ G
by solid (blue), dash-dotted (green), and dashed (red) curves, 
respectively. This parameter is 
independent of ion charge and mass numbers. Dotted vertical lines
of the same colours show the densities at which the ground Landau level
is completely filled for $Z/A=1/2$ and $Z/A=1/5$ for the same magnetic 
fields (these fractions along with respective magnetic fields 
are written near the lines). For realistic 
materials, $Z/A \approx 1/2$ and therefore, our consideration is valid
quantitatively at densities higher than the left-most dotted line
for each field. 
This corresponds to $b \lesssim 1$, i.e. moderate and weak magnetization
of crystal phonons.    

As already mentioned, the case of electrons on the ground Landau level,
and thus $b \gg 1$, requires a separate consideration.

Additionally, at sufficiently low temperatures one has to take into 
account bandgaps in the electron spectrum of the 
crystal \citep[see][for details]{RY82,C12}, whereas we use the
free-electron approximation.

\section{Conclusion}
We have calculated electrical and thermal conductivities of degenerate
electrons emitting and absorbing phonons in a strongly magnetized 
crystalline neutron star crust. The novel features of our study include 
{\it (i)} an account of modification of the phonon properties in the 
magnetic field and {\it (ii)} an application of a generalized 
variational method to solve the Boltzmann kinetic equation in the 
magnetic field with due consideration of the electron-phonon process 
inelasticity.

Our results apply at weak and intermediate phonon 
magnetization $b = \omega_{\rm B} / \omega_{\rm p} \lesssim 1$ and 
indicate a significant growth of electrical and thermal resistivities 
at quantum and intermediate temperatures $T \lesssim 0.1 T_{\rm p}$ 
in a broad range of chemical compositions and mass densities of matter. 
This effect is due to an appearance of a soft phonon mode in the 
magnetized ion Coulomb crystal, which is easier to excite in an 
interaction with an electron than an ordinary acoustic phonon in the 
field-free case.  

At $b \gg 1$ our results hint at an even stronger increase of 
the electrical and thermal resistivities, however, a detailed analysis 
of the problem is needed under the assumption that electrons populate 
only the ground Landau level. 

These results are important for quantitative modelling of  
cooling of neutron stars and evolution of their magnetic 
fields. In particular, they imply an accelerated decay of the magnetic 
field in the outer neutron star crust accompanied by an additional heat
release.

\section*{Acknowledgments}
The author is grateful to Prof.\ D.G.\ Yakovlev for reading the 
manuscript and making useful remarks. This work was supported 
by RSF, grant No.\ 14-12-00316.

\end{document}